\documentclass[sigconf]{aamas} 

\usepackage{balance} 
\usepackage{amsmath, amsfonts}
\usepackage{tikz}
\usetikzlibrary{arrows.meta}
\usetikzlibrary{shapes.geometric}
\usepackage{clrscode3e}
\usepackage{algorithm}

\setcopyright{ifaamas}
\acmConference[AAMAS '23]{Proc.\@ of the 22nd International Conference
on Autonomous Agents and Multiagent Systems (AAMAS 2023)}{May 29 -- June 2, 2023}
{London, United Kingdom}{A.~Ricci, W.~Yeoh, N.~Agmon, B.~An (eds.)}
\copyrightyear{2023}
\acmYear{2023}
\acmDOI{}
\acmPrice{}
\acmISBN{}

\acmSubmissionID{332}

\title{Learning Parameterized Families of Games}

\author{Madelyn Gatchel}
\affiliation{
  \institution{University of Michigan}
  \city{Ann Arbor}
  \country{USA}
}
\email{gatchel@umich.edu}

\author{Bryce Wiedenbeck}
\affiliation{
  \institution{Davidson College}
  \city{Davidson}
  \country{USA}
}
\email{brwiedenbeck@davidson.edu}

\begin{abstract}
Nearly all simulation-based games have environment parameters that affect incentives in the interaction but are not explicitly incorporated into the game model.
To understand the impact of these parameters on strategic incentives, typical game-theoretic analysis involves selecting a small set of representative values, and constructing and analyzing separate game models for each value.
We introduce a novel technique to learn a single model representing a family of closely related games that differ in the number of symmetric players or other ordinal environment parameters. 
Prior work trains a multi-headed neural network to output mixed-strategy deviation payoffs, which can be used to compute symmetric $\varepsilon$-Nash equilibria. 
We extend this work by making environment parameters into input dimensions of the regressor, enabling a single model to learn patterns which generalize across the parameter space.
For continuous and discrete parameters, our results show that these generalized models outperform existing approaches, achieving better accuracy with far less data.
This technique makes thorough analysis of the parameter space more tractable, and promotes analyses that capture relationships between parameters and incentives.
\end{abstract}


\newcommand{\BibTeX}{\rm B\kern-.05em{\sc i\kern-.025em b}\kern-.08em\TeX}
\newcommand{\abs}[1]{\left| #1 \right|}
\newcommand{\dotproduct}{\kern-.06em \boldsymbol{\cdot} \kern-.06em}

\begin{document}


\pagestyle{fancy}
\fancyhead{}


\maketitle 


\section{Introduction}
Game-theoretic analysis uses mathematical models to investigate incentives in multi-agent systems.
Like all forms of mathematical modeling, this relies on numerous simplifying assumptions, but the constraints that arise from describing utilities and computing equilibria uniquely complicate game-theoretic modeling and analysis. 
In particular, the unobservability of utilities and out-of-equilibrium counterfactuals contributes to a preference in the field for seeking generalized insight from abstract games rather than placing too much trust in precisely tailored numerical models.
One consequence of this imbalance is that game theory has, to its detriment, thus far derived less benefit than other fields from advances in data science, simulation, and machine learning.
In this paper, we introduce a machine-learning-based approach to game-theoretic analysis that helps to bridge the gap between theoretical and numerical methods.

In pencil-and-paper approaches to game theory, researchers often analyze abstract classes of games and describe general properties of their equilibria.
In contrast, existing computational tools typically focus on representing and solving a particular numerical instantiation, what we call a \emph{game instance}, and may therefore struggle to demonstrate robust results or provide adequate comparative analyses.
Our new approach uses a neural network to represent a parameterized \emph{family} of game instances, enabling a rich variety of computational methods which generalize game-theoretic modeling and analysis across some space of environment parameters. 
By learning parameterized families, we provide a tractable computational model that can capture a greater portion of the expressiveness that would otherwise be the domain of purely theoretical analysis.

This work is motivated by the setting of simulation-based games \citep{EGTA2020} (SBGs), where a multi-agent interaction of interest is modeled via an agent-based simulation, and data from those simulations is used to construct a normal-form game.
In typical SBGs the underlying agent-based model has many free parameters.
Examples include the number of background traders in a study of latency arbitrage by high-frequency traders \cite{latency_arbitrage}, the recovery rate in a model of debt consolidation among financial institutions \cite{portfolio_compression}, or the probability of defaults in a simulation of credit network liquidity \cite{credit_networks}.
In some cases, appropriate values for environment parameters can be chosen by empirical validation, but much more often, several possible settings for a parameter are plausible, and may require deliberate exploration.
In each example above, distinct normal-form game models were constructed from separate sets of simulation data and analyzed independently for multiple values of the given parameter.
Further, all of these simulation-based models have many other parameters of potential strategic relevance, and it seems likely that they were under-explored due to this burden of constructing entirely new SBGs.

Our method builds on the work of Sokota et al. \cite{learn_dp}, who demonstrated that a neural network mapping symmetric mixed strategies to deviation payoffs---the expected utilities for unilateral deviators---could be learned from sampled payoff data and could stand in for a normal-form payoff matrix in computing $\varepsilon$-Nash equilibria. 
Fundamentally, this is made possible by player symmetries in the environment, which are exhibited by nearly all large SBGs.
Our core insight is that this approach can extend far beyond representing a single normal-form game by making other parameters of the studied interaction endogenous to the learned model.
We show that by training a neural network to map symmetric mixed strategies \emph{and environment parameters} to deviation payoffs, we can produce a single model that generalizes over a family of games, which can dramatically reduce the sample complexity of certain comparisons and enables other altogether new analyses.
Our results show that this generalized model achieves better accuracy than existing approaches, using significantly less data.


\section{Related Work}
This paper extends a long line of work on improving methods for analyzing simulation-based games, and connects with a range of studies applying machine learning techniques to game theory.
It also addresses a clear need for better approaches to generalizing across environment parameters in simulation-based games.

\subsection{Simulation-Based Games}
In a typical simulation-based game, an agent-based model is used to capture the dynamics of the environment being studied, and different possible behaviors of the agents in that environment can be simulated \cite{EGTA2020}.\footnote{Also known as empirical games \cite{egta} or black-box games \cite{ef_blackboxgames}.}
An analyst implements a set of strategies that could govern agent behavior, and then a single run of the simulator assigns a strategy to each agent and produces a noisy sample of the payoffs for that particular profile of strategies.
We take as given that the analyst has access to such an agent-based model, or some other function that maps an arbitrary pure-strategy profile to a (possibly noisy) sample of each player's payoff.
Of note, purely observational payoff data does not in general suffice, because our method requires sampling specific profiles on demand.

In principle, these simulations could be used to fill out a normal-form payoff matrix by sampling every possible profile enough times to produce accurate payoff estimates for that matrix cell; this approach was used by early simulation-based game studies \cite{ad_auc_tac}.
However, this exhaustive sampling approach rapidly becomes infeasible as the model grows, because the number of profiles in a symmetric game grows combinatorially, and for any variation in environment parameters, an entirely new payoff matrix must be constructed. 

\subsection{Varying Environment Parameters}
\label{section:var_ply_games}

\subsubsection{Simulation-Based Games}
\label{section:sbgt_related}
Essentially all simulation-based games employ agent-based models where certain parameters of the simulated environment can be varied, and many examples in the literature include variations over some parameter prominently in their analysis. 
For example, \citet{latency_arbitrage} construct an SBG to analyze the effects of latency arbitrage in financial markets. 
The authors analyze the same game with a variable number of background traders (24, 58, 238), but construct three independent model instances via player reduction \cite{DPR} and analyze each one separately because the number of players is a parameter outside the scope of their game model.
For each instance, they find role-symmetric equilibria in the game and then evaluate background-trader surplus and latency arbitrageur profit.
In another study \cite{fcm_vs_cda}, the same authors compare the welfare of traders in frequent call markets versus continuous double auctions. 
They analyze four separate instances, varying the number of agents (8, 14, 42) and the mean-reversion parameter ($\kappa = 0.05, 0.01$).

\citet{portfolio_compression} use multi-agent simulation to model portfolio compression of debt cycles in a financial network as a strategic decision among firms. 
They define the recovery rate, $\alpha$, as ``the fraction of assets an insolvent node is able to recover and use to pay back creditors, with the remaining assets being lost to default."
They construct separate game-theoretic models for $\alpha \in \{0, 0.1, 0.3, 0.5, 0.7, 1.0\}$, and study the effect of the recovery rate on the incentives to compress debt cycles.

Other simulation-based game examples include \citet{protocol_compliance}, who evaluate complex network routing protocols, \citet{credit_networks}, who study the formation of credit networks, and \citet{price_prediction} who explore prices in simultaneous sealed-bid auctions.
\citeauthor{protocol_compliance} analyze separate instances that vary the number of non-attacking nodes (clients, ISPs, roots, and servers).
\citeauthor{credit_networks} analyze instances that vary the probability that a debtor will default, and the availability of information about those defaults.
\citeauthor{price_prediction} vary buyer's valuation distributions and degree of substitutability among the resources being sold.
In all of these cases, the authors had to build separate models from distinct data sets for each game instance.

\subsubsection{Other Variable-Parameter Games}
\label{section:other_ex}

Game-theoretic analyses that generalize over variable environment parameters are also common outside of simulation-based game settings.
\citet{tcp} model TCP sessions with variable cost and between 2 and 10 users.
\citet{election_eq} model voting in plurality elections with between 3 and 96 voters, and find different trends with even or odd numbers of players.
\citet{fatima_auctions} compares sequential and simultaneous auctions by varying the number of objects and bidders.
\citet{rich_smart} study information in Cournot competition and vary the number of ignorant and informed agents.
In all of these examples, the environment parameters are parameters of the model, meaning the authors have to construct separate game instances for each distinct parameter combination they want to consider.
\citet{famous_ijcai2018} study games with varied parameters, but focus on two-player zero-sum games and address parameter uncertainty from the agent’s perspective as opposed to the analyst’s.

\subsection{Methods for Simulation-Based Games}

\label{section:ML_related}

In the vast majority of simulation-based game studies, sampling from the agent-based simulator is the primary bottleneck, so the literature on simulation-based game theory includes a wide range of approaches for improving data efficiency.
Some tools employ statistical methods such as control variates \cite{ad_auc_tac} and bootstrapping \cite{bootstrap} to reduce the number of samples required to confidently estimate a profile's payoffs.
Others focus on iteratively exploring the space of possible strategies \cite{strategy_exploration,credit_networks} to identify equilibria from partial payoff matrices.
Games with a large number of players have been tackled via player-reduction approaches \cite{HR,DPR} which analyze a small game that is hoped to roughly approximate a larger one.
Various attempts \cite{greenwald,bounding_regret} have been made to derive theoretical bounds on sample complexity.
We prioritize the practicality of our approach for SBGs over these sorts of theoretical guarantees that have limited applicability to large and complex game families.

Several machine learning methods have also been proposed for simulation-based games.
Some of these aim to identify compact structure, such as role symmetry \cite{TR}, a graphical game \cite{learning_graphical_games} or both \cite{structure_learning}, underlying a particular data set.
But they can only help in the event that such structure is present but not known in advance.
Others aim to learn models that can replace the payoff matrix data structure, such as \citet{learn_pf_inf_games} who propose regression approaches for learning payoff functions in games with real-valued strategies, and
\citet{fengjun} who use Gaussian process regression to learn the utility function in symmetric games with a large number of players.

Most directly relevant is the work of Sokota et al. \cite{learn_dp}, who use a neural network to learn a mapping from role-symmetric mixed-strategy profiles to deviation payoffs. 
They exploit the player symmetries common in simulation-based settings to learn the deviation payoff function in games with a large number of players. 
This learned deviation payoff function is used in Nash-finding algorithms to identify role-symmetric approximate equilibria in SBGs without constructing an explicit payoff table.
Our approach extends that of Sokota et al. \cite{learn_dp}, adding environment parameters as an input to the network and allowing us to model parameterized game families. 


\section{Background}
A normal-form game consists of a set of players, indexed by $i$, a set of strategies $S_i$ for each player, and a utility function for each player: $u_i : \prod_i S_i \rightarrow \mathbb{R}$.
A game is \emph{symmetric} if any permutation of the player set yields the same game, and \emph{role-symmetric} if there exists a non-trivial partition of the player set into roles where permutations within a role yield the same game.
Our learning technique relies on symmetries, and in this paper we focus on symmetric games, but all our methods also apply under role symmetry.

Exploiting symmetry, we can rewrite the model to be independent of players' identities.
A symmetric normal-form game, $\Gamma$, has a number of players $p$ and one strategy set $S$, indexed by $j$, common to all players.
The utility function shared by all players depends only on an individual's choice of strategy and the aggregate choice of strategies by others,
so it can be represented by a separate payoff function $u_j$ for each strategy $j \in S$ that maps \emph{opponent profiles} to utilities.
An opponent profile $\vec{s}$ is an integer vector of dimension $\abs{S}$ representing the number of other players selecting each strategy.
The set of all opponent profiles $\vec{S}$ is thus an integer simplex:
\[ \vec{S} = \left\{ \vec{s} \in  \mathbb{Z}^{\abs{S}} \;:\; \vec{s}_j \ge 0,\, \textstyle\sum_j \vec{s}_j = p-1 \right\} \]
and the payoff function for each strategy maps this integer simplex to the reals: $u_j : \vec{S} \rightarrow \mathbb{R}$.

The main solution concept used to predict behavior in normal-form games is the \emph{mixed-strategy Nash equilibrium}.
A mixed strategy $\sigma$ is a probability distribution over a player's strategies, and a Nash equilibrium is a mixed strategy for each player where no player can increase their expected utility by unilaterally deviating to any other strategy.
In symmetric and role-symmetric games, analysts are often interested in symmetric or role-symmetric Nash equilibria, where all identical players play the same mixed strategy.
In a symmetric game, a symmetric mixed-strategy profile $\vec{\sigma}$ is a probability distribution over $S$, according to which all $p$ players (or sometimes all $p-1$ opponents) independently randomize their actions.
Note that $\sigma$ and $\vec{\sigma}$ have the same dimension, but the vector-accent emphasizes that in a symmetric profile, many players are following the same mixed strategy.

A symmetric mixture $\vec{\sigma}$ is a Nash equilibrium if a unilateral deviator's expected utility for playing any pure strategy is no higher than their expected utility for playing $\sigma = \vec{\sigma}$, when all opponents play according to $\vec{\sigma}$.
To express this mathematically, we extend our notation for utility functions to map symmetric mixed-strategy profiles to \emph{deviation payoffs}---the expected utility of playing strategy $j$ against opponents playing $\vec{\sigma}$.
This expectation $u_j(\vec{\sigma})$ is a probability-weighted sum of $u_j(\vec{s})$ over all opponent profiles:
\begin{equation} u_j(\vec{\sigma}) = \sum_{\vec{s} \in \vec{S}} \Pr(\vec{s} | \vec{\sigma}) u_j(\vec{s}) \label{eq:devpay}\end{equation}

We define $u: \Delta^{\abs{S}} \rightarrow \mathbb{R}^{\abs{S}}$ (with no subscript) as a function that produces a vector of the deviation payoff for each strategy, given a symmetric mixed-strategy profile $\vec{\sigma}$ from the $\abs{S}$-dimensional probability simplex $\Delta^{\abs{S}}$.
We can then express the Nash equilibrium condition in terms of deviation payoffs:
a symmetric mixed-strategy profile $\vec{\sigma}$ is a Nash equilibrium if $\forall j \in S$, we have $u_j(\vec{\sigma}) \le \sigma \dotproduct u(\vec{\sigma})$, where the dot product of $\sigma$ and $u$ gives the expected utility of playing the same mixed strategy as everyone else.

It is also useful to define \emph{regret}, the maximum gain achievable by deviating from $\vec{\sigma}$ to any strategy $j \in S$, as:
\[ \epsilon(\vec{\sigma}) = \max_{j \in S} \; u_j(\vec{\sigma}) - \sigma \dotproduct u(\vec{\sigma}) \]
A Nash equilibrium has $\epsilon(\vec{\sigma}) = 0$, but in practice, analysts generally identify $\varepsilon$-equilibria---mixtures with $\epsilon(\vec{\sigma}) \le \varepsilon$, for suitably small $\varepsilon$.

\subsection{New Terminology}
The premise of this work is that for settings which vary an environment parameter, we can construct and learn a single game-theoretic model and then perform analyses to characterize trends for the entire game family. 
We capture this idea by first defining an \textit{instance}, which is a normal-form game $\Gamma$ with fixed environment parameters.
Next, a \textit{parameterized game family} is a set of game instances that are related by one or more ordinal parameters of the environment.
Formally, we define a parameterized game family as $\gamma(V) = \{\Gamma(v): v \in \mathbb{R}\}$ for some environment parameter variable $V$. 
Many environments will have multiple variable environment parameters of interest, and our method can simultaneously generalize across all of them, but for our initial validation we focus on parameterized game families with one varied parameter.

A symmetric mixed-strategy profile $\vec{\sigma}$ can be played by any number of symmetric players (regardless of environment parameter values), but because the utility functions vary with the environment parameter $V$, we need to specify $V=v$ to compute regrets and identify equilibria. 
The deviation payoff $u_j(\vec{\sigma}, v)$ for $\Gamma(v) \in \gamma(V)$ is the expected utility of a unilateral deviator playing strategy $j \in S$ while all other opponents randomize according to $\vec{\sigma}$, and is still calculated by equation 1.
In a parameterized game family $\gamma(V)$, the regret $\epsilon(\vec{\sigma}, v)$ of a given mixture, as well as other functions of equilibrium, varies as a function of $V$, and as a result, a mixture may be an $\varepsilon$-Nash equilibrium for certain values of $V$ and not others.


\begin{figure}[t]
    \centering
    \includegraphics[width=2.75in]{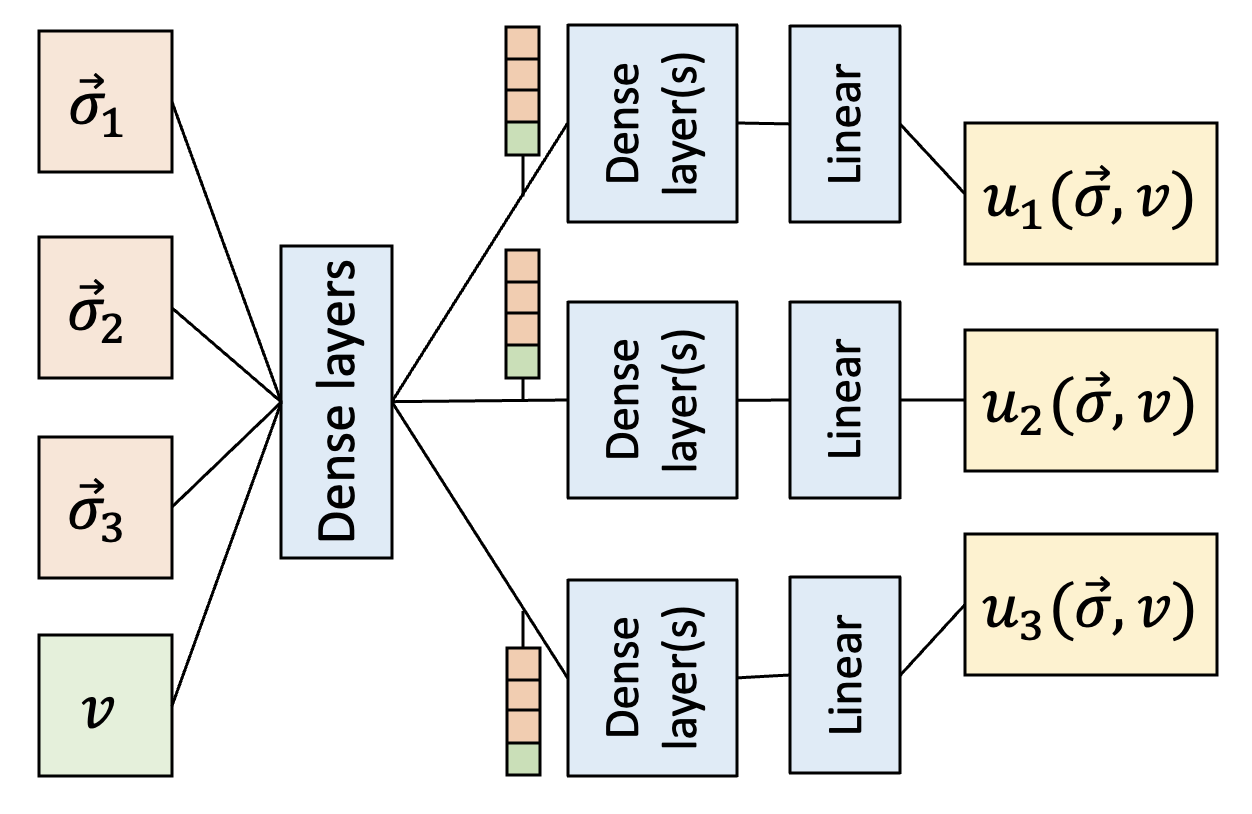}
    \caption{The neural net for variable-parameter learning (VPL) maps symmetric mixtures $\vec{\sigma}$ and environment parameter values $V=v$ to deviation payoff vectors $u(\vec{\sigma}, v).$} 
    \label{fig:vpl_architecture}
\end{figure}

\section{Learning Game Families}
\label{section:model}
We hypothesize that game instances which are related by a common environment parameter likely have related payoff and deviation payoff functions. 
With this hypothesis, we learn a mapping from symmetric mixed strategies and environment parameter $V$ to deviation payoffs in a parameterized game family.
Our approach generalizes that of Sokota et al. \cite{learn_dp} by adding the environment parameter as an input dimension of the model. 
This enables a \textit{single model} to learn patterns which generalize across the parameter space.
In Section~\ref{section:appx_dp} we describe how to construct, train and refine this variable-parameter learning model. 

\subsection{Approximating Deviation Payoffs} 
\label{section:appx_dp}
We use data from an agent-based simulator or other black-box game model to train a neural network on the mapping $u(\vec{\sigma}, v)$ from a symmetric mixed-strategy profile $\vec{\sigma}$ and environment parameter $v$ to the vector of deviation payoffs $[ u_j(\vec{\sigma}, v) : j \in S ]$.
We adapt the technique developed by Sokota et al. \cite{learn_dp} by adding the environment parameter $V$ as an input dimension.
Additionally, in the variable-parameter learning (VPL) model each strategy head has a skip connection from the input layer.
We verified experimentally (not shown) that these skip connections slightly improved VPL performance.
Figure~\ref{fig:vpl_architecture} shows a general architecture for our variable-parameter neural network on a hypothetical game with 3 strategies.
Note that $\vec{\sigma}_j$ denotes the probability that the $j$th strategy is played according to $\vec{\sigma}$ and that  $u_j(\vec{\sigma}, v)$ denotes the deviation payoff of strategy $j$ in instance $V=v$, the expected utility a player receives by playing strategy $j$ when all opponents follow $\vec{\sigma}$.
Observe the added input dimension for the environment parameter and the input layer skip connections for each strategy head. 

The variable-parameter learning procedure is described in Algorithm~\ref{alg:ne_alg}.
For input samples, we draw mixtures $\vec{\sigma} \sim \mathrm{Dir}( \vec{\alpha})$ from a Dirichlet distribution with $\vec{\alpha} < \mathbf{1}$; this helps to ensure accurate estimates for small-support profiles.
To generate the associated environment parameter values, we randomly select values $v$ uniformly across the entire parameter space.
For neighborhood sampling, near each returned candidate Nash equilibrium $\vec{\sigma}^*$ with associated environment parameter value $v^*$, we draw neighborhood mixtures $\vec{\sigma}' \sim \mathrm{Dir}(\vec{\alpha} = \omega_{\vec{\sigma}} \cdot \vec{\sigma}^* + 1)$ where $\omega_{\vec{\sigma}} >> 1$.
We draw \textit{normalized} neighborhood environment parameter values $v'$ from a Beta (or other) distribution centered around $v^*$.

\begin{algorithm}[t]
\begin{codebox}
\Procname{\proc{AppxNashEq}($\gamma(V)$, initQueries, resampQueries, numIters):}
\zi $([\vec{\sigma}], [v]) \leftarrow \text{initSamples}(\gamma(V), \text{initQueries})$
\zi $[\vec{s}] \leftarrow \operatorname{sampleOppProfiles}([\vec{\sigma}], [v])$
\zi $[\vec{u}] \leftarrow \text{simulator}([\vec{s}], [v])$ 
\zi data $\leftarrow ([\vec{\sigma}], [v], [\vec{u}])$
\zi regressor.fit(data)
\zi \Repeat 
    \zi ($[\vec{\sigma}^*], [v^*]) \leftarrow \operatorname{findNash(regressor)}$ // parallelized
    \zi $([\vec{\sigma}'], [v']) \leftarrow \text{sampleNbhd}([\vec{\sigma}^*], [v^*], \text{resampQueries})$
    \zi $[\vec{s}'] \leftarrow \text{sampleOppProfiles}([\vec{\sigma}], [v])$
    \zi $[\vec{u}'] \leftarrow \text{simulator}([\vec{s}'], [v'])$
    \zi data $\leftarrow \text{data} + ([\vec{\sigma}'],[v'], [\vec{u}'])$
    \zi regressor.fit(data)
\zi \Until numIters
\zi \Return ($[\vec{\sigma}^*], [v^*]) \leftarrow \operatorname{findNash(regressor)}$
\end{codebox}
\caption{Finding $\varepsilon$-equilibria for a parameterized game family.}
\label{alg:ne_alg}
\end{algorithm}

\subsubsection{Targeted resampling in parameter space}
\citet{learn_dp} demonstrated the importance of targeted resampling in the strategy-space neighborhood of candidate equilibria.
Because of our hypothesis that neighboring game instances have similar payoff functions (and therefore equilibria), we also conduct targeted resampling in the neighborhood of the associated environment parameter value.

\subsubsection{Avoiding Duplicate Queries} 
Unlike past approaches to sim-ulation-based game analysis \cite{learn_dp,fengjun}, we avoid deliberately repeating queries to the simulator on the same profile.
While this would increase the accuracy of payoff estimates \textit{for that profile}, we find that---given a limited sampling budget and the tiny fraction of mixtures being sampled in a typical game---additional queries are better allocated for a wider range of profiles.

\subsubsection{Intermediate Regret Check} 
Because regret varies as a function of parameter $V$, a symmetric mixed strategy may be an approximate equilibrium in one instance but not others. 
As a result, we propose two modifications during the refinement process: filtering out returned candidate Nash that fall below some intermediate regret threshold for the associated instance, and having no intermediate regret filter. 
The intuition is that we do not want to use too many more queries for areas of the simplex which are less likely to contain $\varepsilon$-Nash equilibria.
However, we also not want to have too strict of an intermediate regret filter such that the refinement process favors some game instances more than others.
We evaluate these two algorithm variants in Section~\ref{section:model_refinement}.

\subsubsection{Parallelized Nash-Finding} 
After the model is trained, we run a Nash-finding algorithm using the learned deviation payoff function. 
This Nash-finding algorithm is run in parallel by using the neural network's batch processing to compute deviation payoffs for many mixtures at once, even for different game instances. 
Given a matrix whose columns correspond to $(\vec{\sigma}, v)$ vectors, the neural network can output a matrix of deviation payoff estimates in time proportional to the size of the network as opposed to the size of the exponentially larger underlying normal-form game family. 
This matrix of deviation payoffs is then used to update the mixtures, also in parallel, as specified by the particular Nash-finding algorithm.
The Nash-finding algorithm returns a matrix of candidate equilibria and associated parameter values. 
To our knowledge, this is the first algorithm in the literature that can compute approximate Nash equilibria in parallel for an entire game family. 

\begin{center}
\begin{figure*}[h]
\begin{tabular}{ccc}
\includegraphics[scale=0.25]{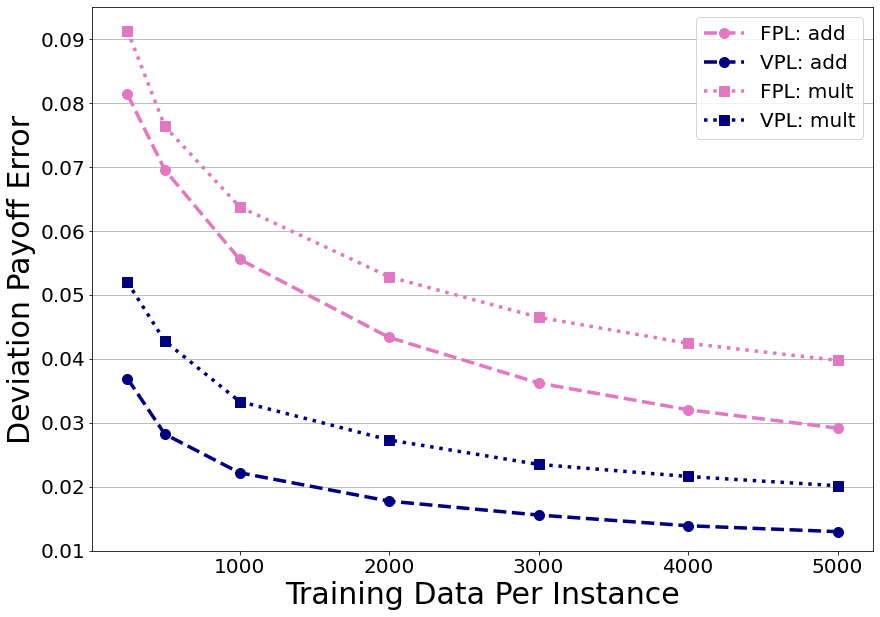} &&
\includegraphics[scale=0.25]{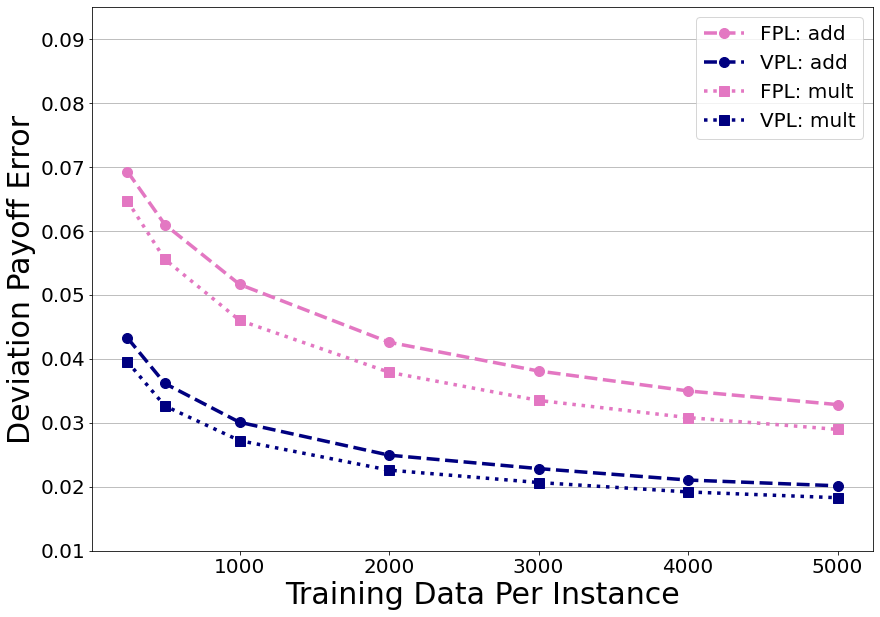} \\
(a) \textbf{Number of Players:} 90-100 && (b) \textbf{ER Threshold:} 0.15-0.25
\end{tabular}
\caption{For both of the most difficult cases: (a) variable number of players and (b) a continuous environment parameter, a single Variable-Parameter Learning (VPL) model outperforms a collection of Fixed-Parameter Learning (FPL) \cite{learn_dp} models on both classes of random games given identical training data.
This suggests that the VPL model is more data efficient, utilizing payoff information from neighboring game instances to approximate deviation payoffs with greater accuracy.}
\label{fig:fpl_vs_vpl_var_ply}
\end{figure*}
\end{center}

\subsubsection{Analyzing the Learned Model}
After the last iteration, the analyst can use the newly refined model to construct a set of candidate equilibria for the parameterized game family $\gamma(V)$. 
First, the analyst could run the Nash-finding algorithm using the learned deviation payoff function to find approximate equilibria for each instance (as described above). 
Then they could apply any of the techniques proposed in Section~\ref{section:analysis} that use this set of candidate equilibria to perform analysis that generalizes across the parameter space.


\section{Experiments}

\label{section:experiments}
The goal of our experiments is to validate that our variable-parameter learning (VPL) model can adequately learn the deviation payoff function for parameterized families of games.
Since our method is the first to handle an entire game family, and prior work \cite{learn_dp} is the current best approach for learning a single instance, our primary experiments compare our VPL model against a collection of fixed-parameter learning (FPL) models.
The next experiment validates our approach to adapting the model refinement from \cite{learn_dp} to the variable-parameter setting. 
Additional experiments with wider ranges of player counts (e.g., 50-100) and Erd\H{o}s-Renyi threshold values (e.g., 0-1) confirm that VPL models continue to perform well, even as the parameter range width increases. 

To serve as proxy for simulator data in all experiments, we generate random symmetric additive (or multiplicative) polynomial-sine bipartite action-graph games with additive function nodes (BAGGFN) as used in prior literature \cite{learn_dp, fengjun}.
These random games have complex but learnable payoff functions, particularly compared to common game distributions in related literature: substantially more challenging than congestion games, but much more structured than uniform random games.
Refer to Appendices~\ref{section:random_game_generation} and ~\ref{section:RPS} for more details on random game generation and justification.

Our experiments vary two key parameters of these types of games, chosen to represent some of the most relevant and challenging simulation-based game settings.
First, we vary the number of players in the game.
Because FPL and all previous methods are based around approximating a normal-form game, the number of players determines the dimension of the underlying data, meaning that changing the number of players always requires starting over from scratch with zero carry-over of data.
Second, we vary a continuous parameter: the Erd\H{o}s-Renyi probability threshold used to determine edge inclusion in the underlying bipartite action-function graph.

\subsection{Comparison to Existing Work}
\label{section:comp_existing_work}
Because the number of simulator queries is most often the bottleneck in SBG studies, we first use \textit{identical training sets} when comparing VPL and FPL techniques.
In particular, we compare a single VPL model trained on the entire parameter space against the performance of several FPL models, one trained for each game instance. 
Thus the VPL model trains on the entire training set and the training dataset is partitioned so that each FPL model gets the appropriate data per instance.
We evaluate the performance of the two techniques measured by the deviation payoff accuracy relative to the underlying game family.
We assess model performance on 100 games from two classes: additive and multiplicative polynomial-sine games, each with 5 strategies. 
For a given game class, both models are evaluated on the same 100 games with identical training sets for each game. 
For example, if the total amount of data is 55,000 then the VPL model is trained with 55,000 training examples and each of the 11 FPL neural networks is trained with 5,000 training examples.

The FPL architecture is identical to that described in \cite{learn_dp}, and consists of a multi-headed neural network with 128-, 64-, and 32-node dense hidden layers and a head for each strategy with a 16-node dense layer followed by a linear layer. 
The VPL architecture consists of a multi-headed neural network with 256-, 128- and 64- dense hidden layers and a head for each strategy with a 32-node dense layer followed by a linear layer. 
Refer to Figure~\ref{fig:vpl_architecture} to see the general VPL architecture visualization.
The hyperparameters for the two models were optimized separately, and for expediency, hyperparameters were tuned using other random BAGGFN games. 
In a simulation-based game, such tuning would be performed on a hold-out set, which is validated in \cite{learn_dp}. 
All 11 FPL neural networks use identical hyperparameters, but tuning them independently yields negligible improvement.

For each instance, we measure network performance on 495 mixtures corresponding to points on a lattice that evenly covers the simplex.
We evaluate both models on all 11 instances of the parameterized game family. 
For each mixture we compute the error between the predicted deviation payoff and the ground truth deviation payoff for each strategy, and compute the mean absolute error across the 5 strategies and average across mixtures.
These errors are computed on normalized deviation payoffs, so the average MAE tells us approximately what percentage we can expect our learned deviation payoffs to differ from the calculated deviation payoffs for each strategy. 
For all experiments we compute 95\% confidence intervals for the deviation payoff error for each instance. For each model, we use the mean of MAEs from 495 mixtures for all 100 randomly generated games. 
Note that many of the confidence intervals are too small to show up in the plot.

Figure~\ref{fig:fpl_vs_vpl_var_ply}a shows FPL and VPL performance on additive and multiplicative polynomial-sine games with 5 function nodes, an Erd\H{o}s-Renyi probability threshold of 0.2, and 90 to 100 players. 
Observe that increasing the training data size by the same amount for both models results in a greater increase in accuracy for FPL, but VPL still achieves better overall accuracy. 
This suggests that VPL is able to utilize neighboring instance payoff data to better approximate deviation payoffs, even with a smaller amount of training data.

Figure~\ref{fig:fpl_vs_vpl_var_ply}b shows FPL and VPL performance on additive and multiplicative polynomial-sine games with 50 function nodes, 100 players, and an Erd\H{o}s-Renyi probability threshold ranging from 0.15 to 0.25.
The increase from 5 to 50 function nodes in these random games results in 10 times as many possible edges compared to the games from Sokota et al. Even with this increased complexity and greater variation among neighboring game instances, our VPL method is able to generalize well across the parameter space.

Not surprisingly, for both varied parameters as the amount of data per instance increases, the deviation payoff error for both models decreases. 
VPL clearly outperforms FPL on both classes of games given the exact same training data, and FPL's performance doesn't approach VPL's until the data-per-instance available to FPL approaches the \emph{total} amount of data used by VPL.
These results suggest that a VPL model is more sample efficient than a collection of FPL models, because it can also make use of data from neighboring instances to improve learning in areas of the simplex for which it has fewer samples.
In Appendix~\ref{section:per_instance_validation}, Figure~\ref{fig:fpl_vs_vpl_per_ply} validates that this trend is consistent across instances.
Based on these results, we conclude our variable-parameter learning model is the better approach to learning the deviation payoff function for a parameterized family of games.

\begin{center}
\begin{figure}[t]
\includegraphics[scale=0.24]{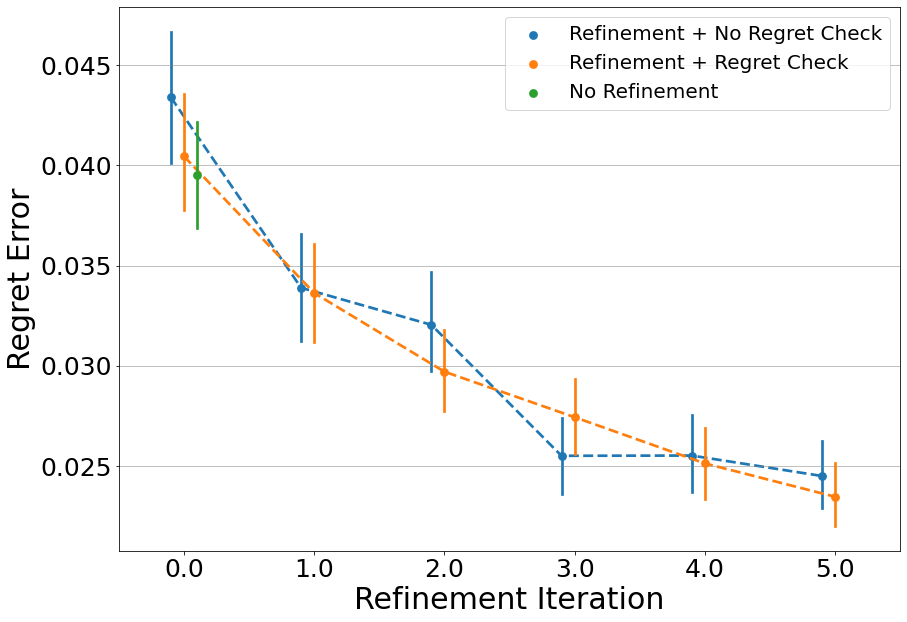}
\caption{Partitioning the training data across several iterations of targeted model refinement reduces approximate equilibrium regret error compared to no model refinement with the same total amount of training data.}
\label{fig:model_refine}
\end{figure}
\end{center}

\subsection{Model Refinement Validation}
\label{section:model_refinement}
Sokota et al. \cite{learn_dp} demonstrated the importance of targeted resampling in the neighborhood of candidate equilibria as a supplement to deviation payoff learning.
Model refinement for parameterized game families also involves refinement over the space of the variable parameter, which is a non-trivial extension given that regret varies as a function of $v$, so an approximate equilibrium in one game instance may not be an approximate equilibrium in others.
We consider two variations of refinement: one where the mixtures from the Nash-finding algorithm are filtered by a maximum regret threshold (as estimated by the model) prior to neighborhood sampling and another where all resulting mixtures are included.
The number of neighborhood samples for each candidate equilibrium is adjusted accordingly (higher for filtered mixtures) to ensure that both refined models train on approximately the same amount of data as the model with no refinement. 

We examine whether and how model refinement can improve the accuracy of equilibrium computations across the parameter space by measuring the absolute error of regret estimates for our reported equilibrium candidates.
We compare the two refined VPL model variants to a model without refinement which trains on the same total amount of data, but gathered all at once rather than in refinement stages.
For this experiment, we use replicator dynamics \cite{rep_dyn} as the Nash-finding algorithm.
We run a large number of replicator dynamics updates in parallel from different initial mixtures for a fixed number of iterations.

For this experiment, the three models---no refinement, refinement with intermediate regret check, refinement with no regret check---are evaluated on the same 100 additive sine games with 5 strategies and 50 to 100 players.
For each refinement iteration we run replicator dynamics on 100 random mixtures per instance (5100 mixtures total). 
All three models are trained on approximately 55,000 payoff samples each. 
The model without refinement is given all training data at once; the two refined models are given 44,000 initial payoff samples. 
One refined model uses a maximum regret of 0.1 to filter approximate equilibria, with 200 neighborhood samples per equilibrium. 
The other refined model does not apply a maximum regret filter, and generates 100 neighborhood samples per distinct mixture resulting from replicator dynamics. 
The regret error is measured as the mean absolute error of regret, normalized to the payoff scale, for all candidate equilibria across all instances for each iteration.
As in previous experiments, we compute 95\% confidence intervals around the average regret error.

Figure~\ref{fig:model_refine} shows that each iteration of targeted model refinement reduces approximate equilibrium regret error, and both refinement approaches significantly outperform the model with the same total training data but no refinement. 
In Appendix~\ref{section:per_instance_validation}, Figure~\ref{fig:model_refine_per_ply} validates that this trend is consistent across instances.
With no clear difference between the two refinement variations, we cannot draw any conclusions about the effectiveness of the intermediate regret threshold. 
We believe an analyst with more specific domain knowledge may be able to make an appropriate choice about whether to include a maximum regret filter for refinement.

\begin{center}
\begin{figure}[h]
    \centering
    \begin{tabular}{c}
        \includegraphics[scale=0.22]{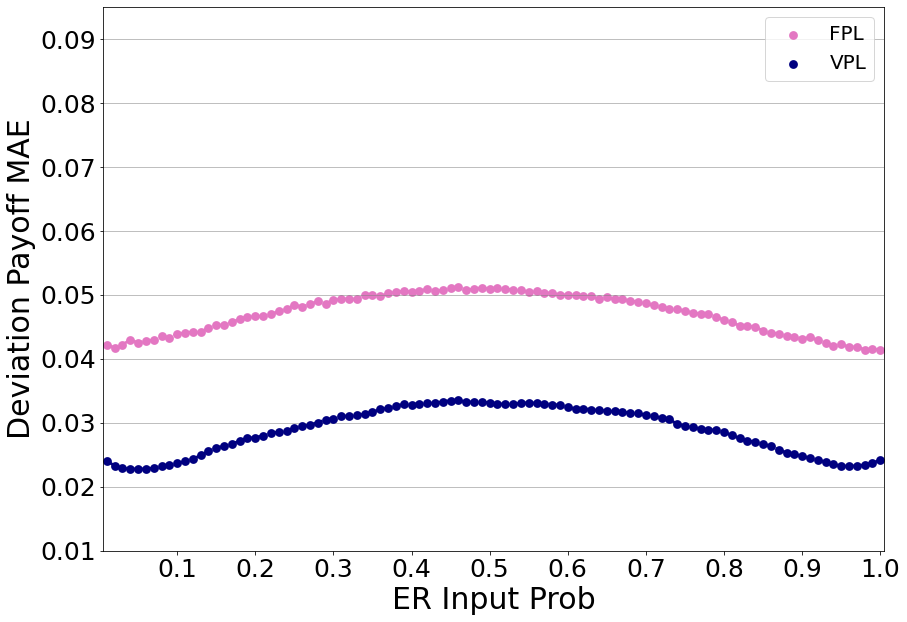} \\
        (a)  \\
        \includegraphics[scale=0.22]{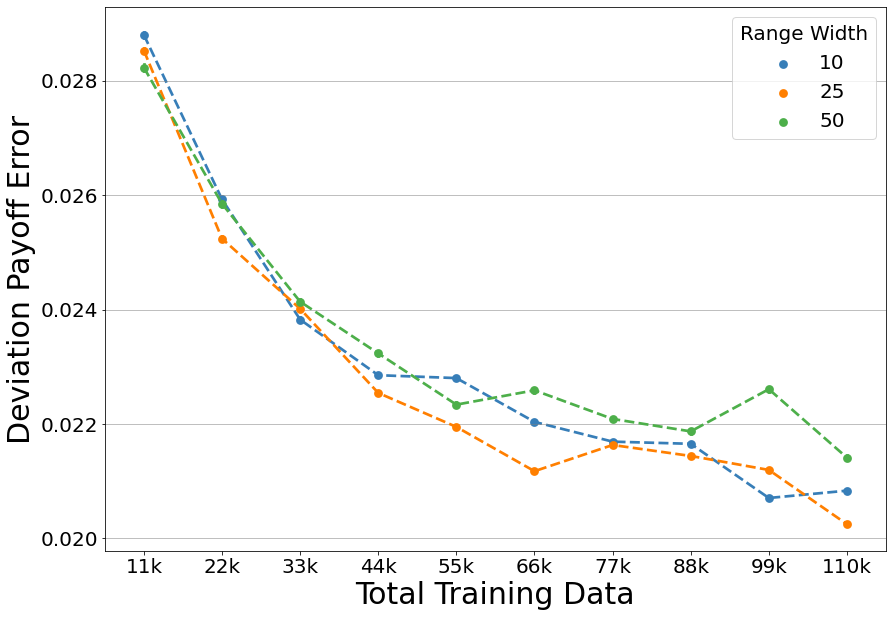}  \\
        (b) 
    \end{tabular}
    \caption{(a) VPL still outperforms FPL on random games with 100 function nodes and with Erd\H{o}s-Renyi threshold varied from 0 to 1. (b) For the three models with different player-count range widths, the deviation payoff errors are roughly the same. }
    \label{fig:scalability}
\end{figure}
\end{center}

\subsection{VPL Scalability Validation} 
\label{section:scalability}
\subsubsection{Variable Erd\H{o}s-Rényi Threshold}
We assess VPL model scalability on random additive polynomial-sine game families with an Ed\H{o}s-Rényi threshold ranging from 0.01 to 1.
These random games have 5 strategies, 100 players, and 100 function nodes --- the most complex random games out of all our experiments.
Figure~\ref{fig:scalability}a compares a single VPL model against 100 FPL models (one per game instance) when trained on 100k total training examples. 
Not only does the VPL model still outperform the collection of FPL models, but the overall VPL deviation payoff error across the entire range is comparable to the best overall VPL deviation payoff error in Figure~\ref{fig:fpl_vs_vpl_var_ply}b where the games had half as many function nodes!

\subsubsection{Variable Number of Players} 
To further evaluate the scalability of our model for discrete parameters we investigate whether increasing the width of the parameter range  affects the deviation payoff error, given the same amount of total training data. 
In this experiment we evaluate VPL models with player-count range widths equal to 10, 25 and 50 on the same set of 100 random 5-strategy, 5-function additive polynomial-sine games for different amounts of total training data. 
To account for varied deviation payoff function complexities associated with different magnitudes of player counts, we evaluate each model's performance by averaging deviation payoff error on game instances with $50, 60, \dots, 100$ players.
This means that we evaluate one VPL model with range width equal to 50 ($50 \leq p \leq 100$ players), two VPL models with range width equal to 25 ($50 \leq p \leq 75$ and $75 \leq p \leq 100$ players), and three VPL models with range width equal to 10 ($50 \leq p \leq 60$, $70 \leq p \leq 80$, and $90 \leq p \leq 100$ players); each VPL model receives the same amount of training data. 
We compute the average normalized deviation payoff MAE across all strategies for 495 mixtures per instance, per game. 
Once again the magnitude of the confidence intervals is too small to see in the plot.

Figure~\ref{fig:scalability}b shows that increasing the total amount of training data improves performance regardless of parameter range size. 
Further, our results show no notable difference in performance between the models with differing player-count range widths, given the same amount of training data.
This suggests that for wider parameter ranges, a single VPL model is still the better approach to learning the deviation payoff function for a parameterized family of games. 
Additionally, VPL is appealing for wider parameter ranges because it is far less tedious to train and refine a single neural network compared to 51 different FPL neural networks, and far more informative than just two snapshots at, say, 50 and 100 players.


\section{Parameterized Game Family Analysis}
\label{section:analysis}
A limitation of work that studies a small number of separate game instances is that the analysis may not adequately depict trends for the entire parameter space. 
As a result, the relationship between the environment parameter and incentives in the game often remains understudied. 
Our single learned model allows for more tractable and complete analysis of the parameter space, and also enables several new types of analysis. 
We now present two possible avenues to help characterize the relationship between environment parameters and incentives.
We emphasize that these are just a glimpse into the many new varieties of analysis that variable-parameter models will enable.

\begin{figure}[t]
    \centering
    \begin{tabular}{cc}
    \includegraphics[scale=0.11]{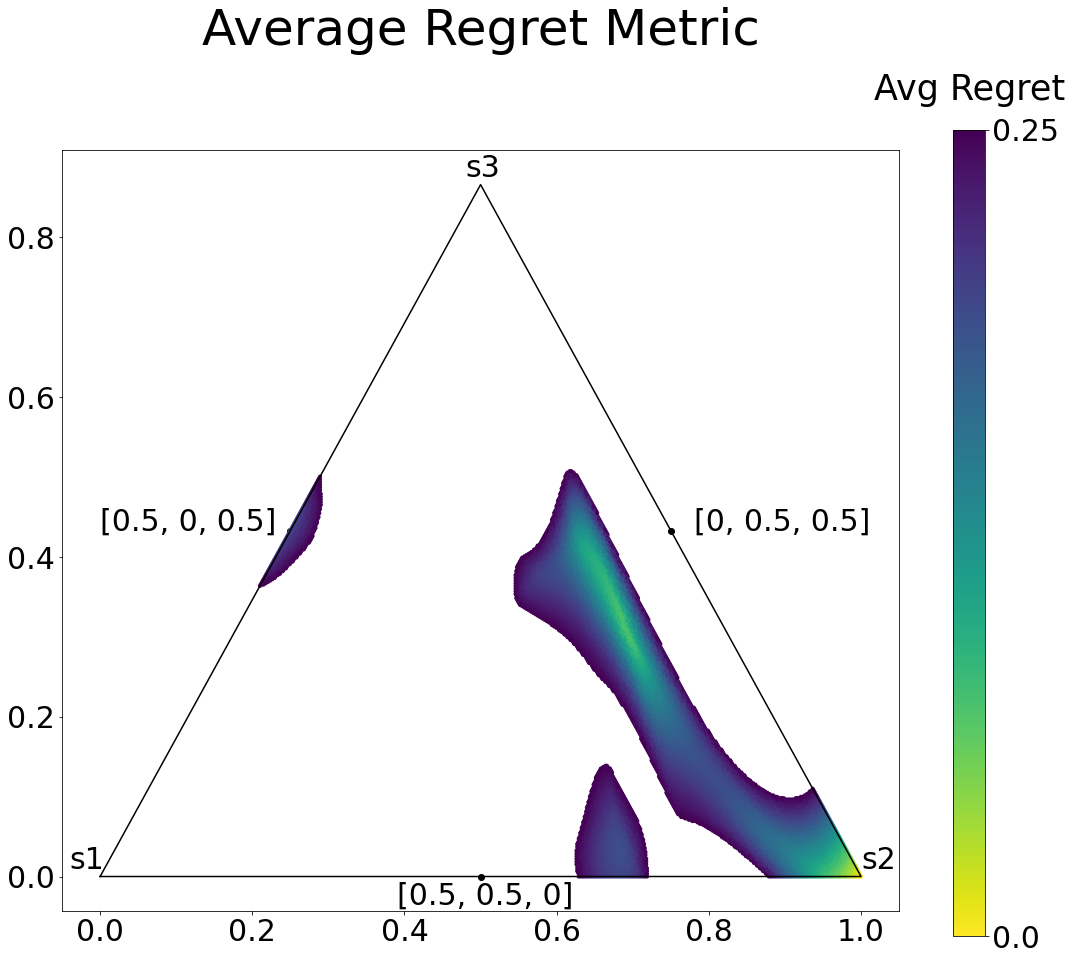} &   
    \includegraphics[scale=0.11]{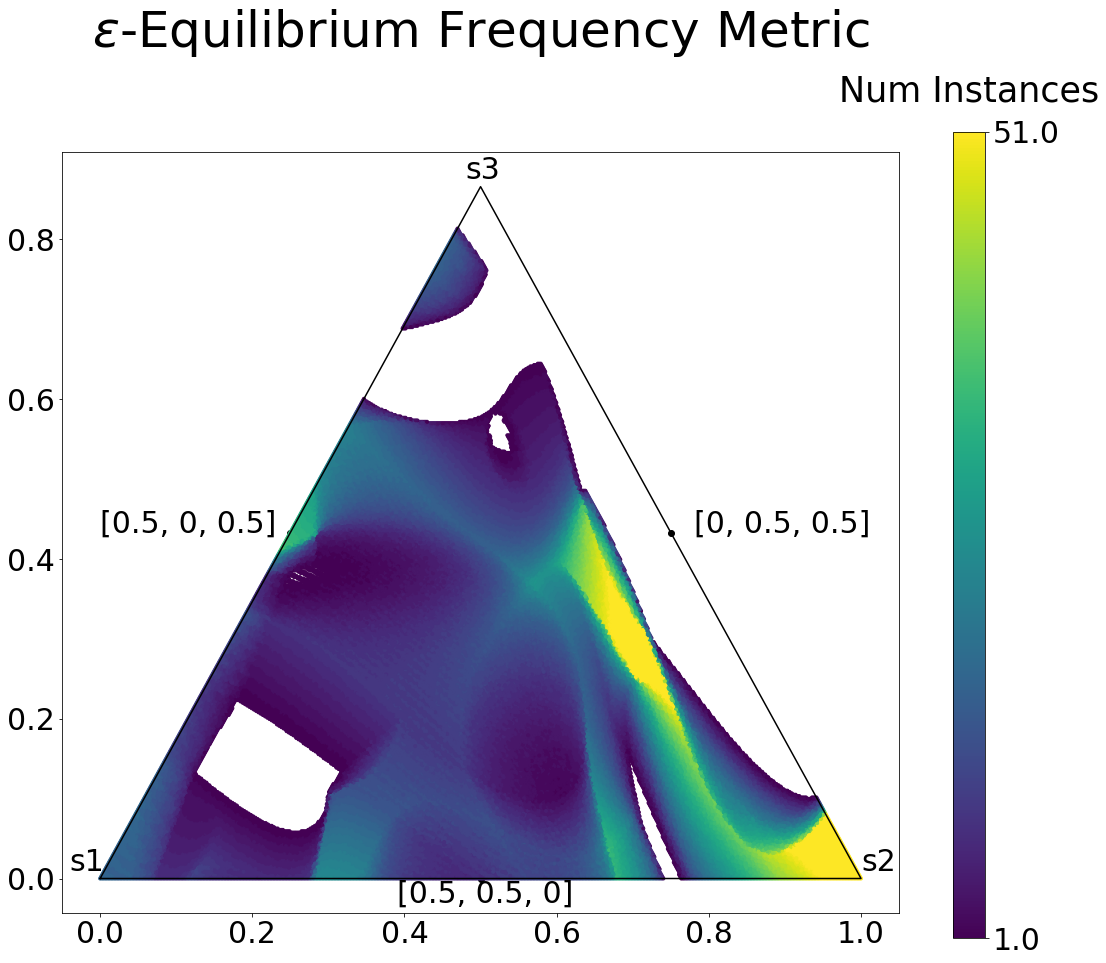} \\ 
    (a) & (b) \\
    \end{tabular}
    \caption{Comparison of two robustness metrics on a randomly generated 3-strategy game: (a) average regret metric, (b) $\varepsilon$-equilibrium frequency metric.}
    \label{fig:comp_robustness}
\end{figure} 

\subsection{Robust Equilibria}
Because a Nash equilibrium is only defined for a game instance with fixed environment parameters, in a parameterized game family $\gamma(V)$ any function of equilibrium, such as regret, price of anarchy or social welfare, is also parameterized by $V$.
In some cases, an analyst with uncertainty about a parameter's true value might want to identify robust equilibria.
We can compute any statistic of the given function of equilibrium, and use the resulting value as a measure of the robustness of the equilibrium.
For example, an analyst might evaluate average or max regret across all game instances for a given profile, and label the profile as a robust equilibrium if the regret statistic is below some threshold.

\subsubsection{Example} Figure~\ref{fig:comp_robustness} compares robust equilibria found by two robustness metrics in a randomly generated 3-strategy multiplicative polynomial-sine game with 50 to 100 players using calculated regrets.
Each point in the simplex corresponds to a symmetric mixed strategy. 
In Figure~\ref{fig:comp_robustness}a, the color shows the mean regret of the corresponding profile across all game instances.
Note that the white points correspond to profiles in which mean regret was greater than $\varepsilon$. 
In Figure~\ref{fig:comp_robustness}b, the color shows how many times each mixture was an approximate equilibrium (for a fixed $\varepsilon$). 
In this plot, the white points correspond to profiles that were never approximate equilibria. 
In both plots, the brighter points correspond to profiles that are considered more robust. 
The similarities between these plots are typical for variable-parameter games we explored---in our experiments we have found that various robustness metrics tend to identify broadly similar sets of profiles as robust equilibria.

\begin{figure}[th]
    \centering
    \includegraphics[scale=0.24]{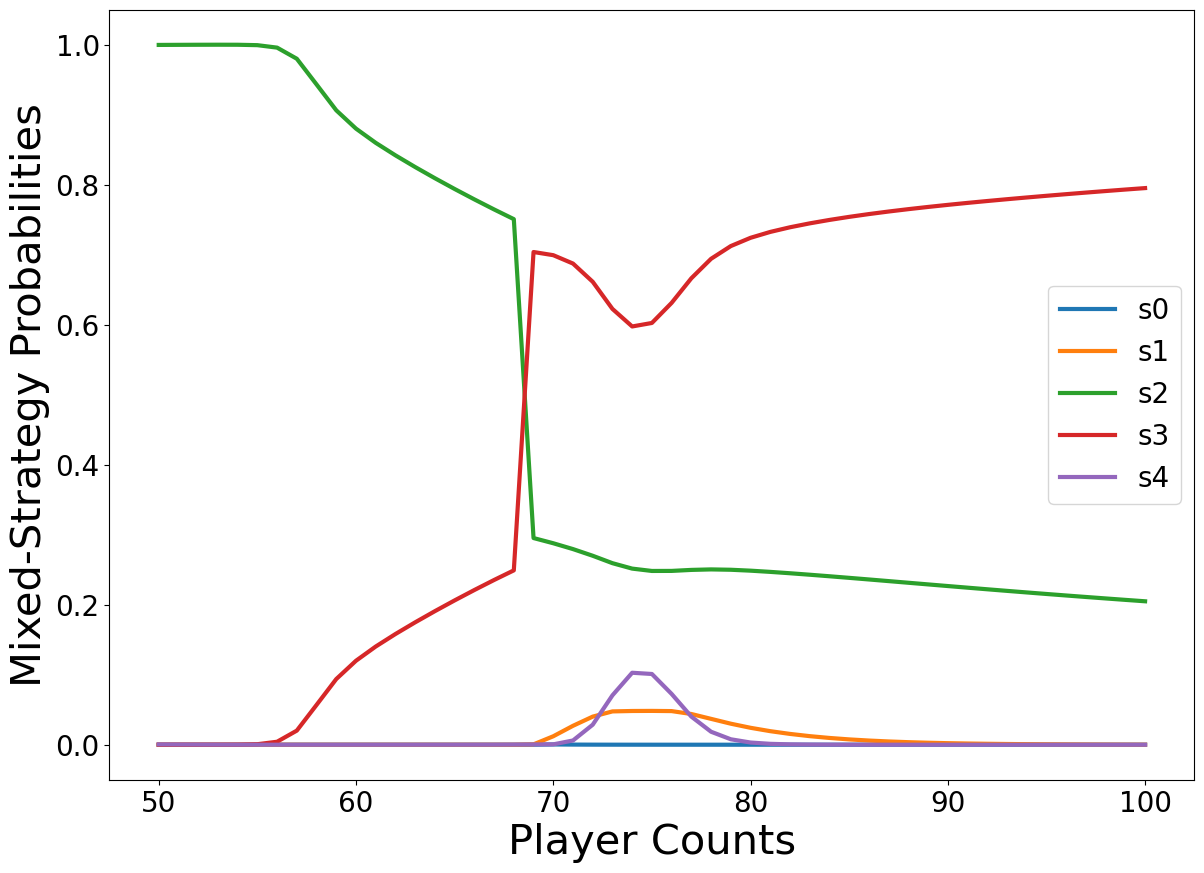}
    \caption{Approximate equilibrium probabilities for each strategy plotted as a function of the number of players for a random multiplicative polynomial-sine game family. This game family demonstrates the importance of parameter sensitivity analysis across the full parameter space to better understand the incentives in the game.}
    \label{fig:nash_sensitivity}
\end{figure}

\subsection{Parameter Sensitivity Analysis} 
For a given parameterized game family $\gamma(V)$, one important question an analyst might want to investigate is: As the parameter $V$ is varied, how do the incentives in the game family change? 
One possible avenue to answer this question is to look at which strategies appear in the support of candidate equilibria across game instances. 
Strategies which appear in the support for many approximate equilibria across game instances may be less sensitive to changes in $V$ and vice versa.
Clustering instances which have (or lack) similar equilibria and then reasoning why the instances are clustered together based on domain-specific knowledge may provide additional insight. 
Finally, identifying equilibrium basins of attraction under a particular search algorithm may help describe how approximate Nash equilibria ``move" in the simplex as the environment parameter is varied.

\subsubsection{Example} Figure~\ref{fig:nash_sensitivity} shows how one $\varepsilon$-equilibrium varies across the player parameter space in a random multiplicative polynomial-sine game family with 5 strategies and 50 to 100 players. 
For each game instance, we ran replicator dynamics with a uniform initial mixture and plotted the equilibrium probabilities for each strategy.
Note that for the entire game family this $\varepsilon$-equilibrium has an average regret of 0.03 and a maximum regret of 0.15. 
If an analyst selects ``representative" player count values 50, 75, and 100 and then constructs and analyzes separate game models for each game instance, they would have an incomplete view of how this $\varepsilon$-equilibrium varies across the parameter space, particularly between game instances with 50 and 75 players. 
Thus this game family demonstrates the importance of analyzing the entire parameter space to better understand an environment's equilibria.


\section{Conclusion}
Our new approach to learning deviation payoff functions that generalize over parameterized game families demonstrates clear advantages over existing techniques for simulation-based games.
Our experiments show that in the common case where analysts want to explore multiple settings for certain environment parameters, learning a single variable-parameter model can produce more accurate analysis from a smaller data set than previous state-of-the-art methods that rely on analyzing each game instance independently.
This variable-parameter learning technique allows for more complete analysis of both continuous and discrete parameter spaces, and enables new types of robustness and sensitivity analysis that previously would have been intractable in simulation-based games.


\begin{acks}
This work was supported in part by funding from the US Army Research Office (MURI grant W911NF-18-1-0208).
\end{acks}


\balance
\bibliographystyle{ACM-Reference-Format} 
\bibliography{references}


\appendix
\newpage
\section{Random Game Generation}
\label{section:random_game_generation}
To serve as a proxy for simulator data, we generate random symmetric games from classes used in prior literature \cite{learn_dp, fengjun}. 
More specifically, we evaluate network performance on randomly generated bipartite action-graph games with additive function nodes (BAGGFN) with 5 strategies.
The game families either vary the number of players from 50 to 100 or vary the Erd\H{o}s-Rényi probability threshold from 0 to 1 (or a subset of those ranges). 
These random games have complex but learnable payoff functions, particularly compared to common game distributions in related literature: substantially more challenging than congestion games, but much more structured than uniform random games.
The compact representation of these random symmetric games allows for efficient ground-truth deviation payoff computation, which is useful to validate the learned models in our experiments. 
Also, the random games can be easily defined with a variable number of players, Erd\H{o}s-Rényi probability threshold, or other environment parameter.

Bipartite action-graph games with additive function nodes (BAGGFNs) are a type of action-graph game \cite{AGGs} where the nodes of the graph can be partitioned into two independent sets, one containing action nodes and the other containing function nodes.
Function nodes are contribution-independent, taking as input the total number of players playing actions in their neighborhood node and outputting the value of their function computed on that neighborhood player-count. 
The payoff for a player playing action $a$ is equal to the weighted sum of the function outputs for the function nodes in the neighborhood of action node $a$. 
In a game with $p$ players, the function table stores outputs associated with 0 to $p$ players.
Thus in a parameterized game family with a variable number of
In a parameterized game family where the number of players ranges from $m$ to $n$, the payoff table associated with the instance with $p$ players, where $m \leq p \leq n$, is a subtable of the payoff table associated with the instance with $n$ players.
This is a particularly nice property of BAGGFNs---to define a BAGGFN with a variable number of players, we generate the game with $n$ players and then all payoff information for instances $m$ to $n$ is already stored in the function table. 

For our experiments, we generate random additive (and multiplicative) polynomial-sine BAGGFNs \cite{learn_dp, fengjun}.
These games derive their name from each function node computing the sum (or product) of a random long-period sine function and a low-degree polynomial with random coefficients. 
The subgraph containing edges from action nodes to function nodes (i.e., function inputs/neighborhoods) is an Erd\H{o}s-Rényi random bipartite graph. 
The associated Erd\H{o}s-Rényi edge threshold is a continuous parameter of the random game, and can be used to easily define a parameterized game family by maintaining which edges are present for different possible values of the threshold.
We refer to this parameter as ``Erd\H{o}s-Rényi threshold" and ``Erd\H{o}s-Rényi input probability" interchangeably.
The subgraph containing edges from function nodes to action nodes is a complete bipartite graph, which means that every function affects every action. 
The action weights are randomly generated according to a normal distribution with mean 0 and standard deviation 1, with some entries randomly masked.

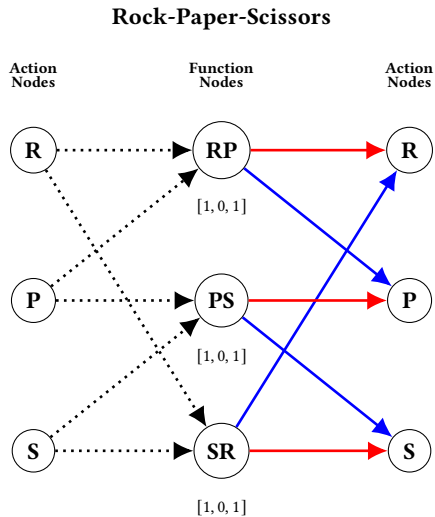
\begin{figure}[htb]
\begin{center}
\textbf{Rock-Paper-Scissors}\\
\vspace{3mm}
\begin{tikzpicture}
\node (aR) [circle, draw] at (0, 0){\textbf{R}};
\node (aP) [circle, draw] at (0, -2){\textbf{P}};
\node (aS) [circle, draw] at (0, -4){\textbf{S}};
\node (RP) [circle, draw] at (2.5, 0){\textbf{RP}};
\node (PS) [circle, draw] at (2.5, -2){\textbf{PS}};
\node (SR) [circle, draw] at (2.5, -4){\textbf{SR}};
\node (R) [circle, draw] at (5, 0){\textbf{R}};
\node (P) [circle, draw] at (5, -2){\textbf{P}};
\node (S) [circle, draw] at (5, -4){\textbf{S}};
\node (FRP) at (2.5, -.75){\scriptsize $[1, 0, 1]$};
\node (FPS) at (2.5, -2.75){\scriptsize $[1, 0, 1]$};
\node (FSR) at (2.5, -4.75){\scriptsize $[1, 0, 1]$};
\node [align=center, text width=1.5cm] (actions1) at (0, 1){\scriptsize \textbf{Action\\[1mm]\vspace{-3mm} Nodes}};
\node [align=center, text width=1.5cm] (func) at (2.5, 1){\scriptsize \textbf{Function\\[1mm]\vspace{-3mm} Nodes}};
\node [align=center, text width=1.5cm] (actions2) at (5, 1){\scriptsize \textbf{Action\\[1mm]\vspace{-3mm} Nodes}};

\foreach \from/\to in {aR/RP, aR/SR, aP/RP, aP/PS, aS/SR, aS/PS}
\draw[dotted, -{Latex[width=2.5mm]}, line width=1pt] (\from) -- (\to);

\foreach \from/\to in {RP/P, PS/S, SR/R}
\draw[-{Latex[width=2.5mm]}, line width=1pt, color=blue] (\from) -- (\to);
\foreach \from/\to in {RP/R, PS/P, SR/S}
\draw[-{Latex[width=2.5mm]}, line width=1pt, color=red] (\from) -- (\to);
\end{tikzpicture}
\caption{RPS represented as a bipartite AGG with additive function nodes (BAGGFN).}
\label{fig:rps_BAGGFNA}
\end{center}
\end{figure}

\section{BAGGFN Example}
\label{section:RPS}
Figure \ref{fig:rps_BAGGFNA} shows Rock-Paper-Scissors (RPS) encoded as a BAGGFN.
We did not use RPS in our experiments; this section is purely included to facilitate understanding about how BAGGFNs work.
There are three action nodes corresponding to actions \textbf{R}, \textbf{P} and \textbf{S}.
There are three function nodes corresponding to ``counters" for the number of players playing (\textbf{R} or \textbf{P}), (\textbf{P} or \textbf{S}), and (\textbf{S} or \textbf{R}). 
The function tables are shown below each function node; all function tables happen to be the same, but this is not usually the case.
Given function table [1, 0, 1] for a given function node $v$, the function node outputs 1 when 0 players are playing strategies in the neighborhood of $v$, 0 when exactly 1 player is playing a strategy in the neighborhood of $v$, and 1 when exactly 2 players are playing a strategy in the neighborhood of $v$.
The blue edges mean that the corresponding edge weight is 1 and the red edges mean that the corresponding edge weight is -1.
We have represented the bipartite graph in this way with action nodes showing up twice so it is easier to see the action nodes passing as input to the function nodes the number of players playing that action and then the function nodes sending the output to the corresponding action nodes.

\begin{center}
\begin{figure*}[t]
\begin{tabular}{ccc}
\includegraphics[scale=0.26]{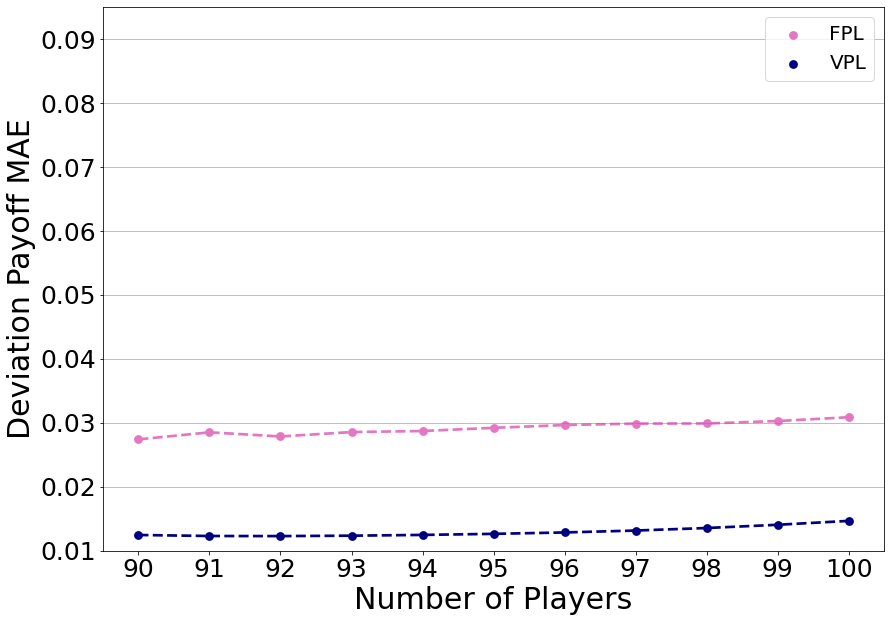} &&
\includegraphics[scale=0.26]{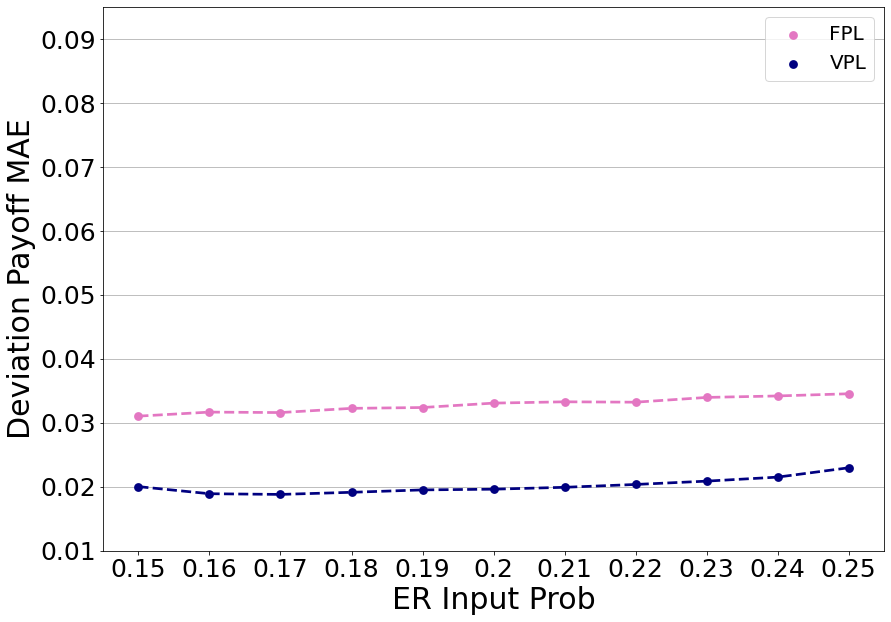} \\
(a) && (b)
\end{tabular}
\caption{Both models perform consistently (i.e., small variation in deviation payoff error) across the (a) discrete parameter space and (b) continuous parameter space, given 5,000 training examples per instance for additive polynomial-sine games.}
\label{fig:fpl_vs_vpl_per_ply}
\end{figure*}
\end{center}

\section{VPL Per-Instance Validation}
\label{section:per_instance_validation}
In our first and second experiments (Sections~\ref{section:comp_existing_work} and \ref{section:model_refinement}), we evaluate how well the VPL model learns the deviation payoff function and how well model refinement improves deviation payoff learning. 
Our experimental results show that VPL significantly outperforms FPL without refinement for both continuous and discrete parameters, and model refinement across the parameter space further improves VPL performance.
In this section, we provide three additional plots which demonstrate that these trends are consistent across all instances.
These plots are generated using the same data from the respective experiments (or some subset thereof), so refer to Sections~\ref{section:comp_existing_work} and \ref{section:model_refinement} for more details on experimental specification and Appendix~\ref{section:hyperparameters} for a summary of relevant hyperparameters. 

Figure~\ref{fig:fpl_vs_vpl_per_ply} compares VPL and FPL deviation payoff performance on additive polynomial-sine games with 5,000 training examples per instance.
The data for these plots is a subset of the data from the corresponding plots in Figure~\ref{fig:fpl_vs_vpl_var_ply} (i.e., per-instance visualization of the right-most additive FPL and VPL points in Figure~\ref{fig:fpl_vs_vpl_per_ply}).
Figure~\ref{fig:fpl_vs_vpl_per_ply}a shows that deviation payoff errors are roughly the same across all game instances when the number of players is varied from 90 to 100. 
Figure~\ref{fig:fpl_vs_vpl_per_ply}b shows that deviation payoff errors are also consistent across all game instances when the Erd\H{o}s-Rényi input prob is varied from 0.15 to 0.25.
Note that the 95\% confidence interval bars are still smaller than the plotted points themselves, even when the y-axis limit width is only 0.01. 
This trend is consistent for all amounts of training data per instance and both classes of games.

\begin{center}
\begin{figure}[t]
\includegraphics[scale=0.25]{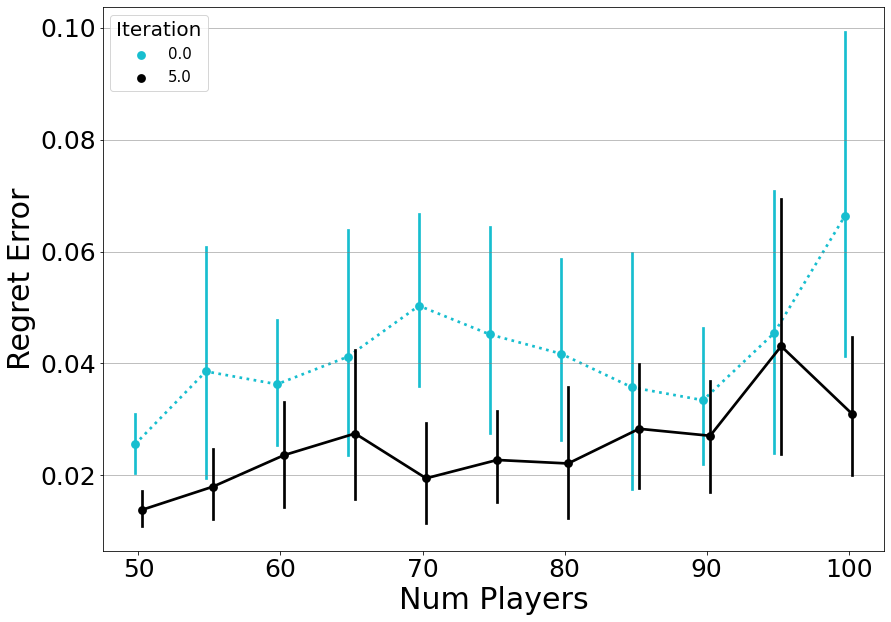}
\caption{Model refinement improves approximate equilibrium regret error roughly evenly across the parameter space.}
\label{fig:model_refine_per_ply}
\end{figure}
\end{center}

Figure~\ref{fig:model_refine_per_ply} shows per-instance Nash equilibrium regret error after the initial training and after 5 iterations of model refinement with 0.1 maximum regret filter and 200 neighborhood samples per approximate equilibrium. 
The results indicate that model refinement improves approximate equilibrium regret absolute error roughly evenly across the parameter space.
The larger 95\% confidence intervals in this plot --- compared with the plots showing deviation payoff errors --- are to be expected. 
This is because there are significantly fewer Nash equilibria per game instance compared to the number of grid mixtures used to evaluate deviation payoff performance.

\begin{center}
\begin{table*}[t]
\begin{tabular}{r|c||c|c|c|c|c}
& Game Type & Per-Inst Train Mix & Dirichlet $\alpha$ & Learning Rate & Batch Size & Num Epochs \\
\hline 
\hline 
Var Players & add & 250, 500, 1k-\textbf{5k} & 0.25, \textbf{0.5}, 0.75, 0.9 & 1e-2, \textbf{1e-3},1e-4 & 16, \textbf{32},100 & 5, \textbf{10}, 15 \\
& mult & 250, 500, 1k-\textbf{5k} &  0.25, \textbf{0.5}, 0.75, 0.9 & 1e-2, \textbf{1e-3},1e-4 & 16, \textbf{32},100 & \textbf{5}, 10, 15 \\
\hline
\hline 
Var ER Thresh & add & 250, 500, 1k-\textbf{5k} &  0.25, \textbf{0.5}, 0.75, 0.9 & 1e-2,\textbf{1e-3},1e-4 & \textbf{16}, 32, 100 & 5, 10, \textbf{15} \\
& mult & 250, 500, 1k-\textbf{5k}  & 0.25, \textbf{0.5}, 0.75, 0.9 & 1e-2,\textbf{1e-3},1e-4 & \textbf{16}, 32, 100 & 5, 10, \textbf{15}                             
\end{tabular}
\caption{Summary of hyperparameters evaluated for the FPL-VPL comparison experiments. Bolded values denote hyperparameters used for final results for both models.}
\label{table:fpl_vpl_hp}
\end{table*}
\end{center}

\section{Summary of Hyperparameters}
\label{section:hyperparameters}
\subsection{Comparison to Existing Work}
Table~\ref{table:fpl_vpl_hp} summarizes the hyperparameters evaluated for the FPL-VPL comparison experiments. 
Bolded values denote hyperparameters used for final results.
Hyperparameters were tuned separately between FPL and VPL models, and bolded values denote hyperparameters used for final results (identical for both models).

Through informal experiments we evaluated the following models on deviation payoff learning performance: linear regressor, random forest regressor, multi-output SVM, and multi-output XGBoost. 
Of these non-neural network models, the multi-output XGBoost model performed the best, but nowhere close to performing as well as a neural network model. 
As a result, we focused our formal architectural tuning on determining whether a multi-headed neural network was needed and whether skip connections from the input layer to each strategy head were helpful. 
We did investigate non-ReLU activation functions in initial experiments but found that ReLU was far superior to others.
Table~\ref{tab:architecture} shows the VPL architectures formally assessed during the validation stage.
Note that `r` denotes ReLU layer, the number denotes the nodes in that layer, `\{r64\}x6` denotes 6 64-node ReLU layers, `H` denotes a strategy head and `S` denotes a skip connection from the input layer. 

\begin{center}
\begin{table}[h]
    \centering
    \begin{tabular}{l|c|c}
    Architecture & M-Head? & Skip? \\
    \hline \hline
    r128;r64;r32;Hr16;Hlin & Y & N \\
    r256;r128;r64;Hr32;Hlin & Y & N \\
    \{r64\}x6;Hr32;Hlin & Y & N \\
    r128;r64;r32;r16;lin & N & N\\
    r256;r128;r64;r32;lin & N & N \\
    \{r64\}x6;r32;lin & N & N \\
    r128;r64;r32;HSr16;HSlin & Y & Y \\
    \textbf{r256;r128;r64;HSr32;HSlin} & Y & Y \\
    \{r64\}x6;HSr32;HSlin &Y & Y \\
    \end{tabular}
    \caption{VPL architectures formally assessed during the validation stage. The bolded architecture is the best-performing architecture and what we used for our test results.}
    \label{tab:architecture}
\end{table}
\end{center}

\subsection{Model Refinement Validation}
For both refined model variants, we separately evaluated the following hyperparameter values: num epochs in \{\textbf{4}, 7, 10\}; num epochs per refinement iteration in \{\textbf{2}, 5\}; $\omega_{\vec{\sigma}} = 100, \textbf{1000}$; and $\omega_v = \textbf{100}, 1000.$
The bolded values denote the values used in the test experiment. 
The model without refinement was trained for 10 epochs.
Additionally, all three models used 100 initial random mixtures per instance for replicator dynamics.

\end{document}